\documentclass[twocolumn,showpacs,fleqn,nobibnotes]{revtex4}

\usepackage{amsmath}
\usepackage{graphicx}
\usepackage{float}
\usepackage{subfigure}

\def\lsim{\raise0.3ex\hbox{$<$\kern-0.75em\raise-1.1ex\hbox{$\sim$}}}
\def\gsim{\raise0.3ex\hbox{$>$\kern-0.75em\raise-1.1ex\hbox{$\sim$}}}

\def\pom{{I\!\!P}}

\newcommand{\rr}{\mbox{\boldmath $r$}}
\newcommand{\rrn}{\mbox{$r$}}

\begin{document}

\title{High energy DVCS on a photon and related meson exclusive production}
\author{Magno V.T. Machado}
\affiliation{Centro de Ci\^encias Exatas e Tecnol\'ogicas, Universidade Federal do Pampa \\
Campus de Bag\'e, Rua Carlos Barbosa. CEP 96400-970. Bag\'e, RS, Brazil}

\begin{abstract}
In this work we estimate the differential cross section for the high energy deeply virtual Compton scattering on a photon target,  $\gamma^* \gamma \rightarrow \gamma \gamma$, within the QCD dipole-dipole scattering formalism. For the phenomenology, a saturation model for the dipole-dipole cross section for two photon scattering is considered. Its robustness is supported by good description of  current accelerator data. In addition, we consider the related exclusive vector meson production processes, $\gamma^* \gamma \rightarrow V\gamma$. This analysis is focused on the light $\rho$ and $\phi$ meson  production, which produce larger cross sections. The phenomenological results are compared with the theoretical calculation using the CD BFKL approach.

\end{abstract}

\maketitle

\section{Introduction}

 It is currently known that at (asymptotic) higher energies, $s\rightarrow \infty$, the parton density inside hadrons increase and the scattering amplitude approaches to the unitarity limit. Therefore, a linear perturbative QCD description  breaks down and one enters the saturation regime, where the dynamics is described by a nonlinear evolution equation and the parton densities saturate \cite{BK,CGC}.  The transition border between the linear and nonlinear regimes is characterized by a typical momentum scale $Q_{\mathrm{sat}}\propto s^{\lambda /2}$, the so-called saturation scale.  The low momentum region is driven by this quantity and these effects might become important in the energy regime probed by current colliders. An interesting process where this approach has been successfully tested is the two virtual photon scattering \cite{Kwien_Motyka}. The formalism considered is based on the dipole picture \cite{dipole}, with the  $\gamma^*\gamma^*$ total cross sections being described  by the interaction of two color dipoles, in which the virtual photons fluctuate into. The dipole-dipole cross section is modeled considering the interpolation between the color transparency domain and the saturation physics. This approach provides a very good description of the data on the $\gamma \gamma$ total cross section for virtual and/or real photons and on the photon structure function at low $x$.

Recently, related studies in the exclusive meson production process $\gamma^*\gamma^*\rightarrow VV$ have appeared \cite{wallon,Enberg,DIvanov}, where $V$ denotes the vector meson.  In Ref. \cite{Enberg}, the BFKL amplitude at next to leading order (NLA) level for the exclusive diffractive two rho production is computed. In Ref. \cite{DIvanov}, the amplitude for the forward electroproduction of two light vector mesons in NLA is computed. A complete program for computation of double diffractive rho production in $\gamma^* \gamma^*$ collisions has been put forward in Ref. \cite{wallon}. A shortcoming of  those approaches is that they consider only the linear regime of the QCD dynamics and  nonlinear effects associated to the saturation physics are disregarded. On the other hand, these processes offer good opportunity for studying the transition between the linear and saturation regimes. Namely, the virtualities of both photons in the initial state can vary as well as the vector mesons in the final state. In the interaction of two highly virtual photons and/or double heavy vector meson production we expect the dominance of hard physics (linear regime). On the contrary, for double light meson production on real photons scattering the soft physics is expected to be important. Consequently, for exclusive processes where we have low momenta scales entering the game we may expect that the main contribution comes from semi-hard physics, described  by saturation effects.

In this letter we study the application of the phenomenological saturation model for two photon scattering to the exclusive hard reactions such  as the amplitude for  deeply virtual Compton scattering (DVCS) on a photon target,
$\gamma^* \gamma \to \gamma \gamma $ and vector meson production, $\gamma^* \gamma \to V \gamma$ ($V=\rho, \,\phi$). The real photon in the final state turns its QCD computation somewhat problematic and saturation model provides a suitable scheme for dealing with low momenta scales appearing in the problem. The present calculation is strongly motivated by recent studies in Ref. \cite{Friot}, where the leading amplitude of the DVCS process on a photon target was considered. Those investigations have shown that the amplitude can be factorized in a convolution of hard processes with the corresponding anomalous generalized parton distributions in the photon, which obey DGLAP-ERBL evolution equations with an inhomogeneous term. Thus, the studies presented here are complementary to those in Ref. \cite{Friot}. In next section, we make a short summary on the dipole-dipole theoretical approach and its phenomenological implementations for two photon scattering. In Sec. 3, estimations of differential cross sections for the purposed processes are presented. Finally, we summarize the results.

\section{Dipole picture for $\gamma\gamma$ scattering}

Let us introduce the main formulas concerning the color dipole picture applied to two-photon scattering. First, we consider the
scattering process $\gamma^*(Q_1) \gamma^*(Q_2) \rightarrow X$, where $Q_i$ stand for the virtuality of the incoming photons. At high energies, the scattering process can be seen
 as a succession on time of three
factorizable subprocesses: i) the photon fluctuates in
quark-antiquark pairs (the color dipoles), ii) these color dipoles interact and, iii) the pairs convert into the virtual photon at final state. Using as kinematic variables the $\gamma^* \gamma^*$ c.m.s. energy
squared $s=W_{\gamma \gamma}^2$, one can define the $x_{12}$ variable \cite{Kwien_Motyka},
\begin{eqnarray}
x_{12}= \frac{Q_1^2 + Q_2^2 + 4m_{q}^2 +  4m_{b}^2}{W_{\gamma \gamma}^2 + Q_1^2 + Q_2^2} \,\,.
\label{bjorken}
\end{eqnarray}
The corresponding imaginary part of the amplitude at zero momentum
transfer reads as \cite{Kwien_Motyka}
\begin{eqnarray}
& & {\cal I}m \, {\cal A}\, (\gamma^* \gamma^* \rightarrow X) =  \sum_{i,j=T,L}\sum_{a,b}^{n_f}
\int dz_1\, d^2\rr_1 \,dz_2\, d^2\rr_2 \nonumber \\
&\times & |\Psi^\gamma_i (z_1,r_1,Q_1)|^2 \,
\sigma_{d d}(x_{12},\rr_1, \rr_2)\,|\Psi^\gamma_j (z_2,r_2,Q_2)|^2 ,
\label{sigmatot}
\end{eqnarray}
where $|\Psi^{\gamma}|^2=(\Psi^{\gamma} \Psi^{\gamma *})$ and $\Psi^{\gamma}_{i}(z,\rr,Q)$  are the light-cone wavefunctions  of the photon
  with polarization $i=T,L$. Photon helicities are implicitly understood. The variable $\rr_1$ defines the relative transverse
separation of the pair (dipole) and $z_1$ $(1-z_1)$ is the
longitudinal momentum fractions of the quark (antiquark) associated to the photon of virtuality $Q_1$. Similar definitions are valid for $\rr_2$ and  $z_2$. Here, $a,b$ stand for quarks of different flavors $a,b=u,d,s,c$. The expressions for the photon wavefunctions are well known \cite{Kwien_Motyka}. The dipole-dipole cross section is represented by the quantity $\sigma_{dd}$, which depends on dipole sizes and on energy through $x_{12}$.

For the case double vector meson
production in the color dipole picture, one consider the
scattering process $\gamma^* \gamma^* \rightarrow V_1 \, V_2$, where $V_i$ stands for
both light and0or heavy mesons. The scattering process can be also seen
 as a succession on time of three
factorizable subprocesses, where now in the step (iii) the pairs convert into the vector mesons. In order to obtain the total cross section, we assume an exponential parameterization for the small $|t|$
behavior of the amplitude. After integration over $|t|$, the total
cross section for double vector meson production by photons is written as,
\begin{eqnarray}
\sigma_{tot}\, (\gamma^* \gamma^*\rightarrow V_1 \, V_2) = \frac{[{\cal I}m \, {\cal A}(s,\,t=t_{min})]^2}{16\pi\,B_{V_1 \,V_2}}\,(1+\beta^{2}) ,
\label{totalcs}
\end{eqnarray}
where $\beta$ is the ratio of real to imaginary part of the
amplitude and $B_{V_1 \, V_2}$ is the slope parameter. The corresponding imaginary part of the amplitude reads as \cite{GMmeson}
\begin{eqnarray}
& & {\cal I}m \, {\cal A}\, (\gamma^* \gamma^* \rightarrow V_1V_2) =  \sum_{i,j=T,L}\sum_{a,b}^{n_f}
\int dz_1\, d^2\rr_1 \,dz_2\, d^2\rr_2 \nonumber \\
&\times & \left(\Psi^\gamma \Psi^{V_1 *}\right)_i \,
\sigma_{d d}(x_{12},\rr_1, \rr_2)\,\left(\Psi^\gamma \Psi^{V_2 *}\right)_j\,
 \, ,
\label{sigmatotmes}
\end{eqnarray}
where now $x_{12}= \frac{Q_1^2 + Q_2^2 + m_{V_1}^2 +  m_{V_2}^2}{W_{\gamma \gamma}^2 + Q_1^2 + Q_2^2}$.
For vector mesons, the light-cone wavefunctions $\Psi^{V_i}(z,\rr)$ are not known
in a systematic way and they should be modeled.  Here, we follows the
analytically simple DGKP approach \cite{dgkp:97}.  The energy dependence of the cross section are shown to be insensitive to the specific model to the meson wavefunction. According model in Ref. \cite{Kwien_Motyka}, for the quark masses one takes $m_{u,d,s}= 0.21$ GeV and $m_{c}=1.3$ GeV.

For phenomenology on dipole-dipole cross section, we take model of Timneanu-Kwiecinski-Motyka (TKM)\cite{Kwien_Motyka}.  The basic idea is that the dipole-dipole cross section  has the same functional form as the dipole-proton one and  is expressed in terms of an effective radius $\rr_{\mathrm{eff}}$, which depends on $\rr_1$ and/or $\rr_2$ \cite{Kwien_Motyka},
\begin{eqnarray}
\sigma_{dd}^{\mathrm{TKM}} (x_{12}, \,\rr_{\mathrm{eff}})  =  \hat{\sigma}_0 \, \left[\, 1- \exp
\left(-\frac{\,Q_{\mathrm{sat}}^2\,\rr^2_{\mathrm{eff}}}{4} \right) \, \right]\,, \label{gbwdippho}
\end{eqnarray}
where $\hat{\sigma}_0 = \frac{2}{3} \sigma_0$, with $\sigma_0=29.12$ mb.  The saturation scale is taken from Ref. \cite{GBW}, where $Q_{\mathrm{sat}}^2=(x_0/x_{12})^{\lambda}$ GeV$^2$, with the parameters $x_0=4.1\cdot 10^{-5}$ and $\lambda = 0.277$.  In Ref. \cite{Kwien_Motyka} three different scenarios for $\rr_{\mathrm{eff}}$ has been considered, with the dipole-dipole cross section presenting in all cases the color transparency property  ($\sigma_{dd} \rightarrow 0$ for $\rr_1 \rightarrow 0$ or $\rr_2 \rightarrow 0$) and saturation ($\sigma_{dd} \rightarrow \hat{\sigma}_0$) for large size dipoles.  Throughout this paper we will consider model I from \cite{Kwien_Motyka}, where  $\rr^2_{\mathrm{eff}} = \rr_1^2 \rr_2^2/(\rr_1^2 + \rr_2^2)$, which is favored by the $\gamma^* \gamma^*$ and $F_2^{\gamma}$ data.

\begin{figure}[t]
\includegraphics[scale=0.47]{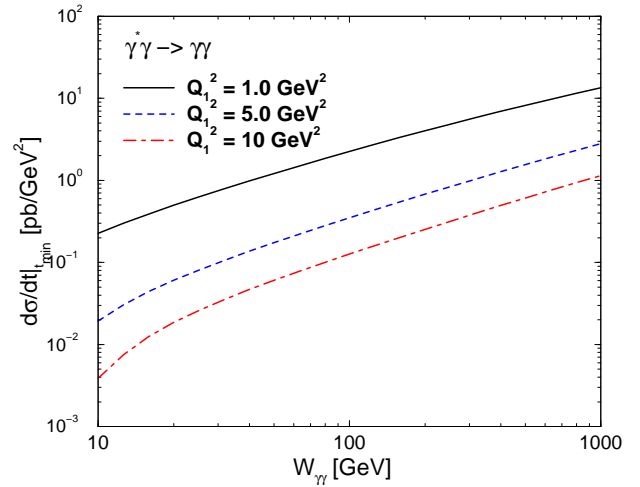}
\caption{(Color online) Differential cross section for DVCS on a photon as a function of energy at fixed photon virtuality  (see text).}
\label{fig:1}
\end{figure}

\section{Results for DVCS on a photon and exclusive meson production}

Let us now introduce the expressions for the differential cross section for the high energy deeply virtual Compton scattering on a photon target within the QCD dipole-dipole scattering formalism. The total cross section is given by $\sigma_{tot}(\gamma^* \gamma \rightarrow \gamma \gamma)= (1/B)\,d\sigma /dt|_{t_{min}}$, where $B$ is the corresponding (unknown) slope parameter. Thus, one has,
\begin{eqnarray}
 \left. \frac{d\sigma}{dt}\right|_{t_{min}}=R_g^2(x_{12},Q^2)\times \frac{[{\cal I}m \, {\cal A}(s,\,t=t_{min})]^2}{16\pi},
 \label{dvcsdiff}
\end{eqnarray}
For the skewedness corrections, coming from the ratio of off-forward to forward gluon distribution, we take parameterization from Ref. \cite{Shuvaev:1999ce}, $R_{g}\,(Q^2)=\frac{2^{2\alpha + 3}}
   {\sqrt{\pi}}\,\frac{\Gamma\,\left(\alpha +
    \frac{5}{2}\right)}{\Gamma \,\left(\alpha +4 \right)}$,
where $\alpha$ is the effective power in energy of the scattering amplitude. For our purpose the amplitude is multiplied by $R_g$, in order to estimate the size of the skewedness effects. This parameterization has been successfully tested in DVCS on a proton  \cite{FM,FMS}.

For now, the real part of amplitude will be disregarded. Due to the specific configuration of DVCS on a photon, only transverse polarization contributes. Thus, the high energy scattering amplitude is given by
\begin{eqnarray}
& & {\cal I}m \, {\cal A}\, (\gamma^* \gamma \rightarrow \gamma \gamma) =  \sum_{a,b}^{n_f}
\int dz_1\, d^2\rr_1 \,dz_2\, d^2\rr_2 \nonumber \\
&\times & \left(\Psi^\gamma \Psi^{\gamma *}\right)_T\,
\sigma_{d d}(x_{12},\rr_1, \rr_2)\,\,|\Psi^\gamma_T (z_2,r_2,Q_2=0)|^2,\nonumber
\label{ampdvcs}
\end{eqnarray}
where
\begin{eqnarray}
\Psi_T^{\gamma}\,\Psi_T^{\gamma\,*} & = &
\frac{6\alpha_{\mathrm{em}}}{4\,\pi^2} \bar{e}_f^2 \, [z^2 +
(1-z)^2] \, \varepsilon_1 \,K_1 (\varepsilon_1 \,\rrn) \,\varepsilon_2
\,K_1 (\varepsilon_2 \,\rrn) \nonumber  \\
 & + &  m_f^2 \, \,K_0(\varepsilon_1\,
\rrn)\,K_0(\varepsilon_2\, \rrn)  \,,
\label{dvcsphoton}
\end{eqnarray}
with $\varepsilon^2_{1}= z(1-z)\,Q_{1}^2 + m_f^2$, $\varepsilon_2=m_f^2$ and $\bar{e}^2_f$ is a summation over the quarks electric charges. In addition,  $|\Psi^\gamma_T|^2$ is the transverse wavefunction squared taken at virtuality $Q^2=0$.

\begin{figure}[t]
\includegraphics[scale=0.47]{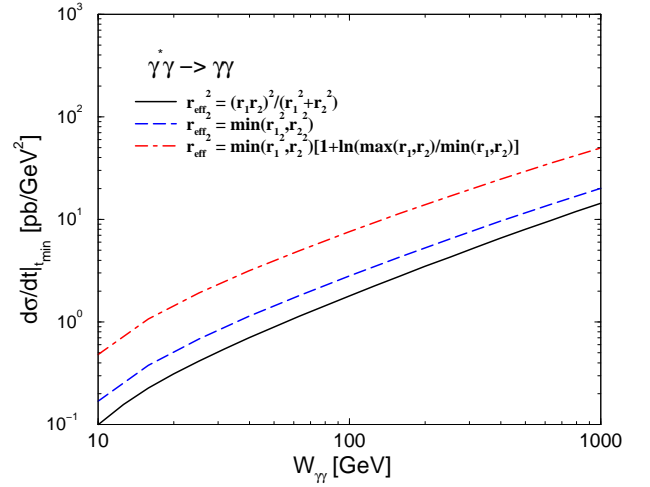}
\caption{(Color online) Comparison of sensitivity on the effective radius model in the dipole-dipole cross section (see text).}
\label{fig:2}
\end{figure}

In Fig. \ref{fig:1} the quantity $d\sigma /dt|_{t_{min}}$ is shown using calculation based on Eq. (\ref{dvcsdiff}). The energy dependence is presented for $W_{\gamma\gamma} \geq 10$ GeV at fixed virtualities $Q_1^2=1,\,5,\,10$ GeV$^2$. In the dipole-dipole cross section we take as the effective radius $\rr^2_{\mathrm{eff}} = \rr_1^2 \rr_2^2/(\rr_1^2 + \rr_2^2)$.
 The effective power of the scattering amplitude ranges on $\alpha = 0.19-0.24$, which increases with $Q^2$. Using a conservative value for the slope parameter, $B\simeq 5.5$ GeV$^{-2}$, based on meson production phenomenology, the total cross section can be evaluated. For instance, at energies larger than 60 GeV one has $\sigma_{tot}(\gamma^* \gamma \rightarrow \gamma \gamma)= 0.12\,\mathrm{pb}\,(W_{\gamma\gamma}/\mathrm{GeV})^{0.77}$ at $Q_1^2=1$ GeV$^2$ and $\sigma_{tot}(\gamma^* \gamma \rightarrow \gamma \gamma)= 3.10^{-4}\,\mathrm{pb}\,(W_{\gamma\gamma}/\mathrm{GeV})^{0.94}$ at $Q_1^2=10$ GeV$^2$. We have tested the sensitivity of the result to a different prescription for the effective radius. We consider the possibilities $\rr^2_{\mathrm{eff}} = \mathrm{min}\,(\rr_1^2,\,\rr_2^2)$ and  $\rr^2_{\mathrm{eff}} = \mathrm{min}\,(\rr_1^2,\,\rr_2^2)[1+\ln \,(\mathrm{max}(\rr_1,\rr_2)/\mathrm{min}(\rr_1,\rr_2)]$ (named model II and III in Ref. \cite{Kwien_Motyka}, respectively).  This comparison is shown in Fig. \ref{fig:2}. It is seen that the model I provides the lower cross section, whereas model III produces values larger by a factor 3. The deviation among the models appears to be independent of virtuality.

Let us now consider the exclusive meson production. In what follows we will take the $\rho$ and $\phi$, which will give larger cross section in contrast with the heavy mesons. Once again, only transverse polarization contributes and the high energy scattering amplitude reads as,
\begin{eqnarray}
& & {\cal I}m \, {\cal A}\, (\gamma^* \gamma \rightarrow V\gamma) =  \int dz_1\, d^2\rr_1 \,dz_2\, d^2\rr_2 \nonumber \\
&\times & \left(\Psi^\gamma \Psi^{V*}\right)_T\,
\sigma_{d d}(x_{12},\rr_1, \rr_2)\,\,|\Psi^\gamma_T (z_2,r_2,Q_2=0)|^2,\nonumber
\label{ampdvcs}
\end{eqnarray}
where now we have to compute the quantity $\Psi_T^{\gamma}\,\Psi_T^{V\,*}$. The
 normalized  light-cone wavefunction for transversely polarized photons is given by:
\begin{eqnarray}
\Psi^{T(\gamma=\pm)} & = & \pm
\sqrt{\frac{N_{c}}{2\pi}} \,e\,e_{f}
 \left[i e^{ \pm i\theta_{r}} (z \delta_{h\pm,\bar{h}\mp} -
(1-z) \delta_{h\mp,\bar{h}\pm}) \partial_{r} \right.\nonumber \\
& + & \left. m_{f} \,\delta_{h\pm,\bar{h}\pm} \right]\frac{K_{0}(\varepsilon r)}{2\pi}\,,
\label{wfT}
\end{eqnarray}
where $\varepsilon^{2} = z(1-z)Q_1^{2} + m_{f}^{2}$.  The electric charge of the quark of flavor $f$ is given by $ee_f$.
The DGKP transverse meson light-cone wavefunction is given by \cite{dgkp:97},

\begin{eqnarray}
& & \Psi_{h,\bar{h}}^{V,T(\gamma = \pm)} = \pm
\left(\frac{i\omega_T^{2}\,r e^{\pm i\theta_{r}}}{m_{V}}\,
[z\delta_{h\pm,\bar{h}\mp} -
(1-z)\delta_{h\mp,\bar{h}\pm}] \right. \nonumber \\
& & + \left. \frac{m_{f}}{m_{V}}\,\delta_{h\pm,\bar{h}\pm}
\right)\,\frac{\sqrt{\pi} f_V}{\sqrt{2 N_{c}}
\,\hat{e}_{f}}f_{T}(z)\,\exp
\left[\frac{-\omega_{L}^{2} \rr^{2}}{2} \right] \,
\label{dgkp_T}
\end{eqnarray}
where $\hat{e}_f$ is the effective charge arising from the sum
over quark flavors in the meson of mass $m_V$. The coupling of the meson to electromagnetic current is labeled
by $f_V^2=3\,m_V\Gamma_{e^+e^-}/4\,\pi\alpha_{em}^2$. The function $f_{T}(z)$ is given by the
Bauer-Stech-Wirbel model \cite{wsb:85}. The meson wavefunctions are constrained by the normalization
condition, which contains the hypothesis that the meson is composed only of
quark-antiquark pairs,  and  by the electronic decay width
$\Gamma_{V\rightarrow e^+e^-}$.

\begin{figure*}[t]
\begin{tabular}{cc}
\includegraphics[scale=0.47]{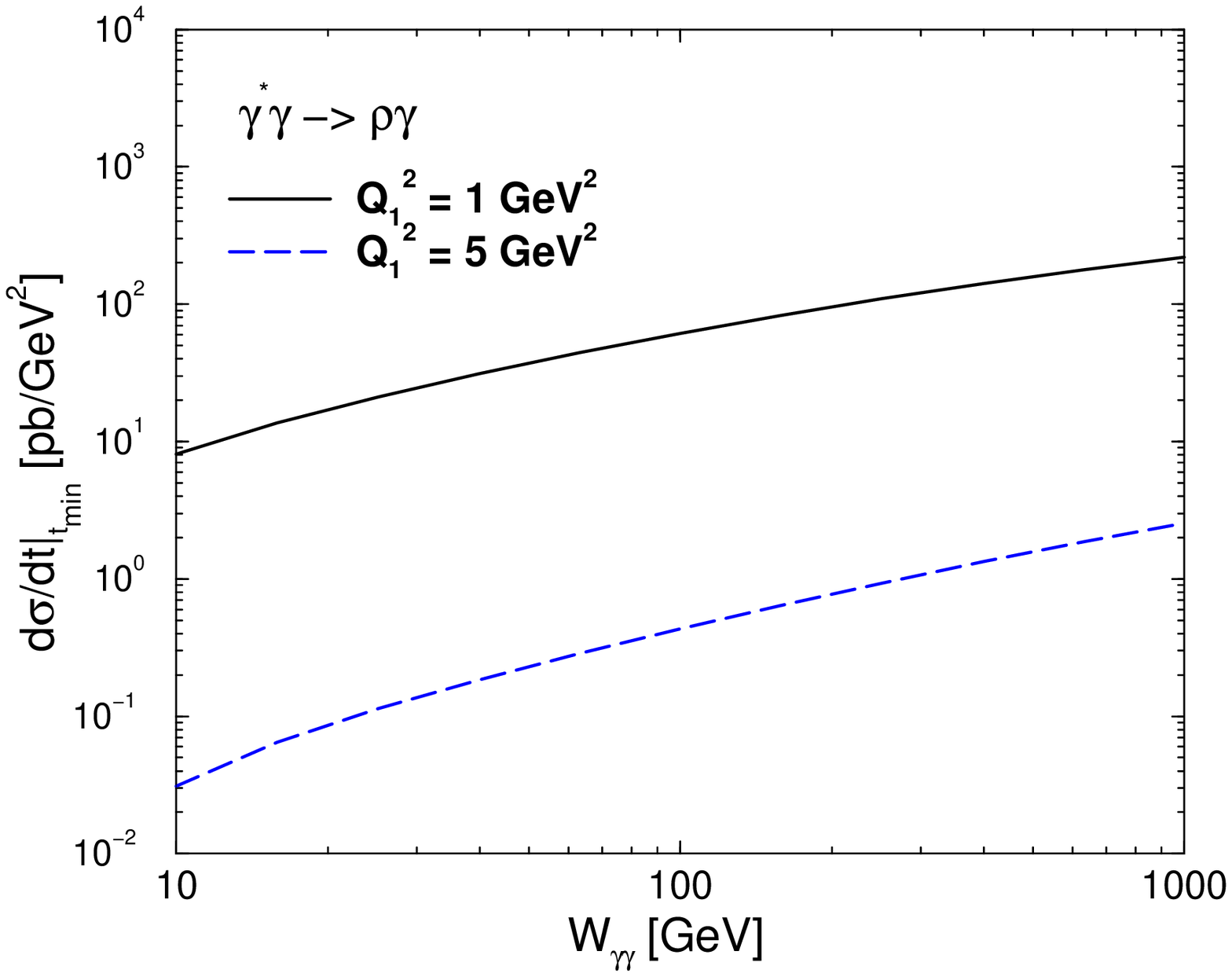} & \includegraphics[scale=0.47]{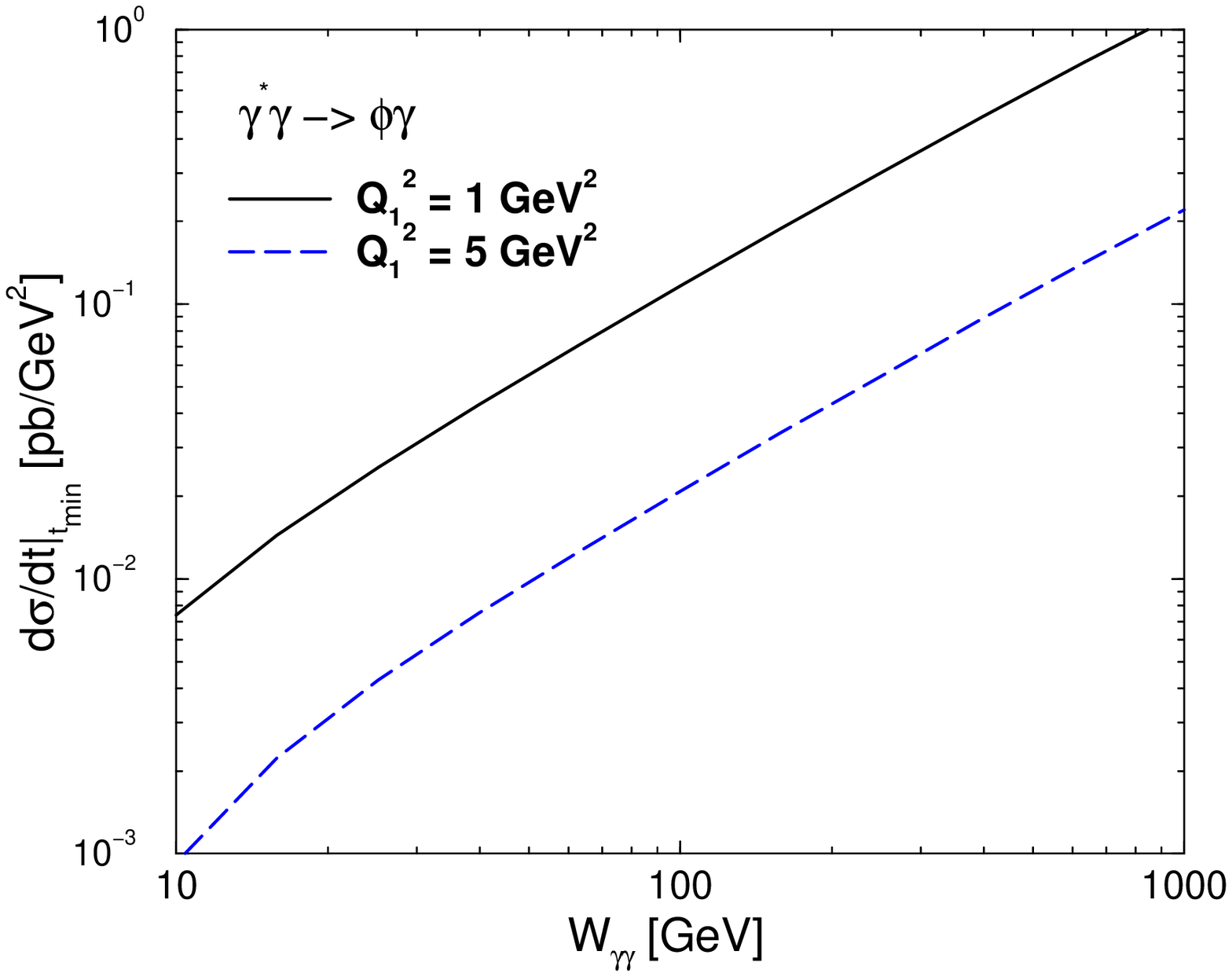}
\end{tabular}
\caption{(Color online) Differential cross section for meson $\rho$ (left panel) and $\phi$ (right panel) exclusive production as a function of energy at fixed virtualities, $Q_1^2=1$ GeV$^2$ and $Q_1^2=5$ GeV$^2$ (see text).}
\label{fig:3}
\end{figure*}

In Fig. \ref{fig:3} (left panel) the quantity $d\sigma /dt|_{t_{min}}$ is shown for the case of $\rho$ production. The energy dependence for values larger than 10 GeV is presented for fixed photon virtualities $Q_1^2=1$ and $5$ GeV$^2$.
 The effective power of the scattering amplitude are $\alpha = 0.14$ and $\alpha=0.19$, respectively. This is consistent with soft pomeron phenomenology. Using $B\simeq 5.5$ GeV$^{-2}$, the total cross section gives $\sigma_{tot}(\gamma^* \gamma \rightarrow \rho \gamma)= 0.6\,\mathrm{pb}\,(W_{\gamma\gamma}/\mathrm{GeV})^{0.56}$ at $Q_1^2=1$ GeV$^2$ and $\sigma_{tot}(\gamma^* \gamma \rightarrow \gamma \gamma)= 2.10^{-3}\mathrm{pb}\,(W_{\gamma\gamma}/\mathrm{GeV})^{0.76}$ at $Q_1^2=5$ GeV$^2$. In Fig. \ref{fig:3} (right panel) the quantity $d\sigma /dt|_{t_{min}}$ is shown for the case of $\phi$ production. The effective power is now $\alpha=0.25$, which is associated with a QCD pomeron due to the larger meson mass. For the total cross section at $Q^2=1$ GeV$^2$ one has $\sigma_{tot}(\gamma^* \gamma \rightarrow \phi \gamma)= 0.5\,\mathrm{pb}\,(W_{\gamma\gamma}/\mathrm{GeV})$.

Here, some remarks are in order. In the calculation above, we have considered the DGKP meson wavefunction. This particular choice presents some shortcomings, as discussed in detail in Ref. \cite{Ivanov}, which contains an extensive discussion of the spin-orbital properties of the wave
function of vector mesons and reports parameterizations consistent with the
S-wave structure of low-lying vector mesons. On the other hand, the DGKP model gives a reasonable phenomenological data description and the results have small sensitivity to the wave function of vector mesons. This can be verified from the recent studies in photoproduction \cite{sandapen,magno_victor_mesons}, where the main effect is a different overall normalization. Thus, a future more detailed study on the model dependence of the results is desirable.

\section{Comparison with other theoretical approaches} 

It is timely to compare the calculation presented here with other theoretical approaches. Although the TKM dipole-dipole cross section is compelling, it is based on phenomenological assumptions. A comparison with the dynamical  Regge-BFKL approach should be valuable. We quote Ref. \cite{cd_bfkl} for complete details on that formalism.

Let us summarize the color dipole BFKL approach. In the color dipole factorization formula the dipole-dipole cross section $\sigma\,(x,r_1,r_2)$ is beam-target symmetric
and universal for all beams and targets, the beam and target dependence is
concentrated in probabilities $|\Psi_A(z_1,r_1)|^{2}$ and
$|\Psi_{B}(z_2,r_2)|^{2}$ to find a color dipole,
$r_1$ and $r_2$ in the beam ($A$) and target ($B$), respectively. If one consider the cross sections averaged over polarizations of the beam and target photons, then only the term $n=0$  of the Fourier series
\begin{eqnarray}
\sigma_{dd}\,(x,r_1,r_2)=\sum_{n=0}^{\infty}\sigma_n(x,r_1,r_2)\,\exp(in\varphi),
\end{eqnarray}
where $\varphi$ is an azimuthal angle  between $r_1$ and $r_2$,
contributes in the dipole factorization formula. It is well known that  that the incorporation of asymptotic 
freedom into the BFKL equation makes the QCD pomeron a series of isolated 
poles in the angular momentum plane. The contribution of the each pole to 
scattering amplitudes satisfies the standard Regge-factorization,
which in the color dipole (CD) basis implies the CD BFKL-Regge expansion for the 
vacuum exchange dipole-dipole cross section
\begin{eqnarray}
\sigma_{dd}\,(x,r_1,r_2)=\sum_{m}C_m\,\sigma_m(r_1)\,\sigma_m(r_2)
\left(\frac{x_0}{x}\right)^{\Delta_m}\,.
\label{eq:2.3}
\end{eqnarray}
Here the dipole cross section $\sigma_m(r)$ is an  eigen-function of the 
CD BFKL equation
\begin{eqnarray}
{\partial\sigma_{m}(x,r)\over \partial \log(1/x)}=
 {\cal K}\otimes \sigma_{m}(x,r)=\Delta_{m}\sigma_{m}(x,r),
\label{eq:2.4}
\end{eqnarray}
with eigen value (intercept) $\Delta_{m}$. For the details on CD formulation of the BFKL equation, infrared regularization by finite propagation radius $R_{c}$ for 
perturbative gluons and freezing of strong coupling at large distances,
the choice of the physically motivated boundary condition for the
hard BFKL evolution and description of eigenfunctions we quote Ref. \cite{cd_bfkl} and references therein. 

Let us point out some features of the quantities presented above. The eigenfunction $\sigma_0(r)$ for the rightmost hard BFKL pole (ground
state) corresponding to the largest intercept $\Delta_0\equiv\Delta_{\pom}$
is node free. The  eigenfunctions $\sigma_m(r)$ for excited states with 
$m$ radial nodes have intercept $\Delta_{m} < \Delta_{\pom}$. The choice $R_c=0.27$fm yields for the rightmost hard BFKL pole the intercept  
$\Delta_{\pom}=0.4$\,, for sub-leading hard poles $ \Delta_m \approx 
{\Delta_0/(m+1)}$. The  node of $\sigma_{1}(r)$ is located at
$r=r_1\simeq 0.05-0.06\,{\rm fm}$, for larger $m$ the rightmost nodes
move to a somewhat  larger $r$ and accumulate at $r\sim 0.1\, {\rm fm}$.  
Because the BFKL equation sums cross sections of
production of multigluon final states, the perturbative two-gluon Born
approximation is a natural boundary condition. This leaves the starting point
$x_{0}$ as the only free parameter which fixes completely the result of
the hard BFKL evolution for dipole-dipole cross section, with the choice $x_0=0.03$. 
 Because in the attainable region of $r$ the sub-leading 
solutions $m\geq 3$ can not be resolved and they all have similar intercepts
$\Delta_m\ll 1$, in practical implementations the expansion (\ref{eq:2.3}) can be truncated at $m=3$ lumping in the term
with $m=3$ contributions of all singularities with $m\geq 3$, with
 with the effective intercept $\Delta_3=0.06$. Whereas scattering of small dipoles $r\lsim R_c$ is  dominated by the 
exchange of perturbative gluons,  interaction of large dipoles with the proton target
 has been modeled by the  non-perturbative
soft pomeron with intercept $\alpha_{\rm soft}(0)-1=\Delta_{\rm soft}=0$. 
Thus, an extra term $\sigma_{\rm soft}(r_1,r_2)$ is added in the  r.h.s. of expansion (\ref{eq:2.3}).

\begin{figure}[t]
\includegraphics[scale=0.47]{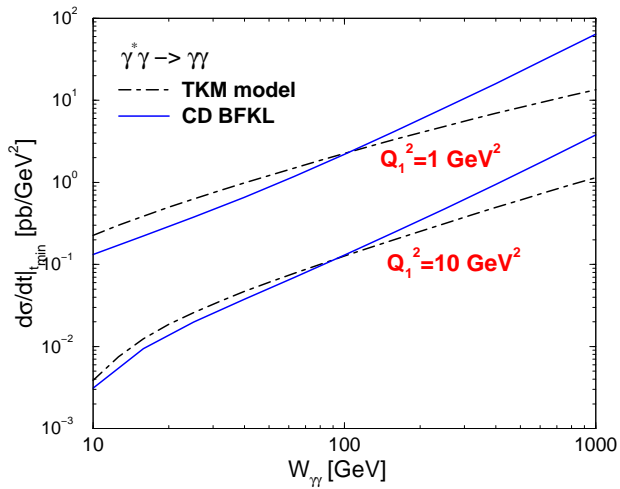}
\caption{(Color online) Comparison between the present phenomenological results with the CD BFKL calculations (see text).}
\label{fig:4}
\end{figure}

In Fig. \ref{fig:4} we compare the TKM model and the color dipole BFKL approach for the dipole-dipole cross section. The cross section $\sigma\,(\gamma^*\gamma \rightarrow \gamma \gamma )$ is computed for two distinct choices, $Q_1^2=1$ GeV$^2$ and $Q_1^2=10$ GeV$^2$. The energy dependence is clearly different at high energies. For sufficiently low energies the results are consistent. The TKM model gives a softer energy dependence in contrast with CD BFKL. This can be understood since at high energies the saturation scale is increasingly larger. The direct consequence of a larger saturation scale is the reduction of the effective energy exponent, $\alpha$. Similar behavior is also observed in vector meson production. Therefore, the high energy limit is a good place to investigate the production dynamics.

\section{Summary}

 We estimate the differential cross section for the high energy deeply virtual Compton scattering on a photon target and also the related exclusive vector meson production processes. The QCD color dipole scattering formalism is considered for phenomenology because it allow to deal with soft momenta scales in problem. The absolute values for cross section are still small, but the exclusive meson production should be feasible. The main theoretical/experimental uncertainties are the slope parameter value and the choice for the meson wavefunction. Skewedness correction have been properly included in calculations, giving a sizable enhancement. We investigated also the model dependence, comparing the present calculations with the results of other theoretical approaches. The CD BFKL formalism is considered and a different energy dependence at high energies is verified.  The present calculation is complementary to the recent QCD approach considering exclusive processes with photon generalized parton distributions. 

\begin{acknowledgments}
This work was financed by the Brazilian funding
agency CNPq. The author is grateful to  Vladimir Zoller (ITEP, Russia) for his valuable help and  for providing the numerical results for the eigen functions and parameters of the CD BFKL approach.
\end{acknowledgments}

\end{document}